\documentclass{article}

\usepackage{PRIMEarxiv}

\usepackage[utf8]{inputenc} 
\usepackage[T1]{fontenc}    
\usepackage{hyperref}       
\usepackage{url}            
\usepackage{booktabs}       
\usepackage{amsfonts}       
\usepackage{nicefrac}       
\usepackage{microtype}      
\usepackage{lipsum}
\usepackage{graphicx}
\usepackage{amsmath}
\usepackage{amssymb}
\usepackage[ruled,vlined]{algorithm2e}

\usepackage{algpseudocode}
\usepackage{subcaption}
\usepackage{algpseudocode}
\graphicspath{{media/}}     
\usepackage[most]{tcolorbox}
\newtcolorbox{inferbox}{
  enhanced,
  colback=blue!3,
  colframe=blue!40!black,
  boxrule=0.6pt,
  arc=2mm,
  left=2mm,right=2mm,top=2mm,bottom=2mm,
}

\newtcolorbox{inferblock}{
  enhanced,
  colback=white,
  colframe=black!25,
  boxrule=0.4pt,
  arc=1mm,
  left=1.5mm,right=1.5mm,top=1mm,bottom=1mm,
}

\providecommand{\kv}[2]{\textbf{#1:}~#2\\}

\title{Knowledge-Driven XRD Phase Identification via
Multi-View Retrieval and Explanation}

\author{
Doaa Mohamed, Markus Stricker \\
Materials Informatics and Data Science\\
Interdisciplinary Centre for Advanced Materials Simulation (ICAMS) \\
Ruhr University Bochum, Bochum, Germany \\
\texttt{\{doaa.mohamed,markus.stricker\}@ruhr-uni-bochum.de}
}

\begin{document}

\maketitle

\begin{abstract}
X-ray diffraction (XRD) is a key experimental technique for determining the phase composition and structure of crystalline materials. However, interpreting XRD patterns is inherently challenging, particularly in high-throughput materials discovery, where many novel materials may need to be characterized and no reference patterns are available for comparison. Consequently, machine learning is increasingly used to accelerate and automate the analysis process while reducing errors associated with human interpretation.

We propose a multi-decision framework for XRD phase analysis that integrates representation learning, similarity-based retrieval, and explainable decision support within a unified reference database. A convolutional autoencoder learns compact latent representations of XRD patterns that preserve structural similarity while remaining robust to variations arising from experimental noise and measurement conditions. By integrating multiple decision pathways within a shared latent space, the framework moves beyond single-label prediction toward ranked and interpretable phase analysis that mirrors expert practice based on reference matching and peak inspection.

During inference, complementary decision mechanisms are applied, including latent-space classification and retrieval, explanation-guided similarity using Integrated Gradients, and composition-based similarity search. These mechanisms generate ranked candidate phase lists that are aggregated into a final prediction with an associated confidence score. Experiments on synthetic datasets demonstrate strong predictive performance, achieving 98.85\,\% accuracy for crystal system classification and 95.82\,\% accuracy for space group prediction on the test set, while maintaining robustness under realistic perturbations. The framework supports reliable, analyst-friendly identification of crystal phases and structures in high-throughput and exploratory materials discovery settings.

\end{abstract}


\section{Introduction}
\begin{figure}[t]
    \centering
    \includegraphics[width=1\linewidth]{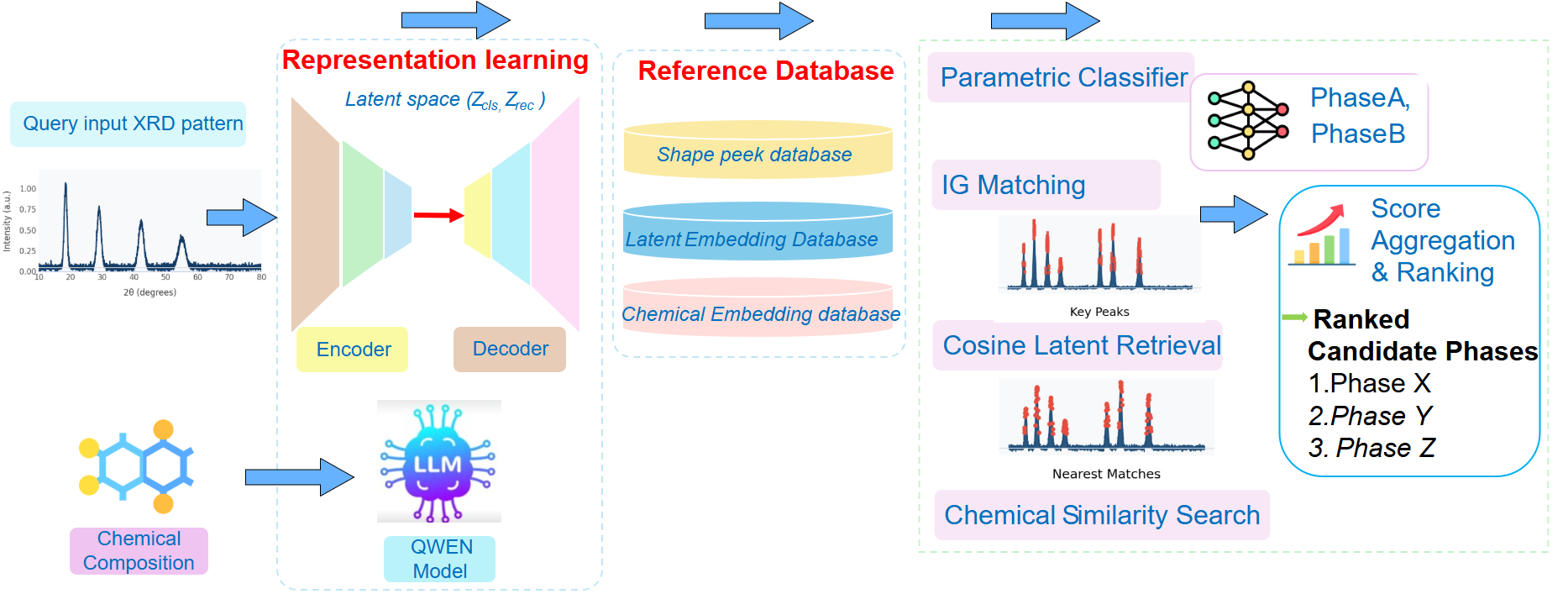}
    \caption{Overview of the HyPhID framework for XRD phase identification. A query diffraction pattern is embedded using a convolutional autoencoder. Predictions combine classifier output, latent-space retrieval, Integrated Gradients (IG) similarity, and composition-based retrieval. The aggregated evidence produces ranked crystallographic predictions.}
    \label{fig:framework}
\end{figure}

High-throughput materials discovery generates large volumes of characterization data that enable the extraction of physical properties and structure-property relationships; however, it requires detailed analyses that are traditionally conducted by human experts~\cite{wang2019rapid,szymanski2021probabilistic}. A prominent example is the vast amount of characterization data produced by X-ray diffraction (XRD), which is a powerful technique for characterizing crystallographic structures, grain size, and molecular structures~\cite{oviedo2019fast}. In an XRD experiment, an X-ray beam interacts with a crystalline sample, producing a diffraction pattern that records scattered intensity as a function of the diffraction angle $2\theta$. The characteristic peaks in this diffractogram encode information about the crystal structure and phase composition of the material. Experimental XRD patterns are typically analyzed by comparing peak descriptors such as positions, intensities, and full widths at half maximum (FWHM) with reference databases such as the Crystallography Open Database (COD)~\cite{Grazulis2012COD} and the Inorganic Crystal Structure Database (ICSD)~\cite{belsky2002icsd}.
Early attempts to automate this process relied on peak indexing~\cite{altomare2009advances} and full-profile
matching algorithms~\cite{gilmore2004high}. However, these approaches often show limited accuracy because experimental spectra may deviate from ideal references due to defects, disorder, impurities, or missing reference patterns for novel materials. More recently, deep learning methods have emerged as powerful alternatives that learn directly from diffraction data and improve robustness to experimental variability~\cite{oviedo2019fast,zaloga2020crystal,vecsei2019neural,maffettone2021crystallography}. Despite their strong performance, most deep learning approaches for XRD analysis remain purely discriminative classifiers that produce single-label predictions, providing limited interpretability and struggling with generalization across synthetic and experimental domains, particularly for structurally ambiguous or previously unseen compounds, which remains an important limitation in the context of materials discovery~\cite{lee2023deep,leeman2024challenges}.\\
Such requirements naturally motivate representation-learning approaches, in which models learn compact latent embeddings that capture the essential structural information contained in diffraction patterns. Variational autoencoders (VAEs) have shown promise for XRD analysis by organizing diffraction patterns into low-dimensional latent spaces where structurally similar patterns cluster naturally~\cite{banko2021deep}. These latent representations enable visualization, clustering, and novelty detection through reconstruction error or latent distance metrics. Beyond visualization and clustering, latent representations also provide a natural foundation for similarity-based retrieval, allowing query patterns to be compared directly against structurally related reference examples.
Interpretability is essential for diffraction analysis because reliable predictions require physically meaningful explanations.
We therefore integrate explainability with similarity-based reasoning to align model decisions with expert-driven peak interpretation workflows.
In addition to interpretability, effective XRD phase identification benefits from flexible decision strategies that combine both parametric and non-parametric reasoning. In this context, parametric models refer to classifiers with learned parameters, such as neural networks, that map diffraction patterns to phase labels through a fixed set of trained weights. These models are efficient and perform well when query patterns closely match the training distribution.  In contrast, non-parametric approaches rely on similarity-based comparison with stored reference examples rather than a fixed parametric mapping. Methods such as nearest-neighbor retrieval in latent representation space allow query patterns to be matched against known reference cases, which can provide more robust behavior in the presence of experimental variability or structural ambiguity~\cite{Wu2018NonParametric,cover1967nearest}.\\
Furthermore, diffraction data alone may not always be sufficient to uniquely identify a phase. Chemical composition provides complementary information that constrains the set of physically plausible structures and can therefore support or refine diffraction-based inference. Recent advances in representation learning enable chemical compositions to be encoded into continuous embedding spaces for similarity-based retrieval~\cite{ward2016magpie,tshitoyan2019unsupervised,zhang2024matnexus}. In contrast, our framework performs decision-level fusion over multiple complementary evidence sources, enabling ranked and interpretable crystallographic inference. These developments have opened new opportunities for incorporating compositional evidence into materials informatics workflows, including phase identification and materials discovery.\\
In this work, we introduce a unified framework for XRD phase analysis that integrates representation learning, similarity-based retrieval, chemical composition information, and explainable decision support to enable phase identification together with the prediction of key crystallographic descriptors, including crystal system and space group. An overview of the proposed framework is shown in Figure~\ref{fig:framework}. In crystallography, identifying the crystal system and space group is a fundamental step in determining the atomic arrangement of a material. The crystal system provides a coarse description of lattice symmetry, while the space group specifies the full symmetry of the structure, making these labels essential descriptors for characterizing and comparing crystalline phases from diffraction data.
A convolutional autoencoder is trained to learn a compact latent representation of augmented experimental and simulated XRD patterns that preserves structural similarity while accounting for variations arising from experimental noise and measurement conditions. After training, the encoder is frozen and used to embed both reference and query patterns into a shared latent space.
A latent reference library is constructed by embedding diffraction patterns generated from crystallographic information files under realistic instrumental, background, and sample-related variations, enabling similarity-based reasoning and interpretable comparison between measured and reference patterns.
For a new diffraction measurement, phase identification is formulated as a multi-decision inference process rather than a single classification step. The query pattern is first encoded using the trained encoder to obtain its latent embedding. Cosine similarity between the normalized query embedding and the latent reference library enables retrieval of structurally similar candidate phases.
In parallel, the classifier provides a prediction over the crystal systems and space groups. Peak-level explanations are computed using Integrated Gradients (IG) to identify influential diffraction regions and support explanation-guided peak matching. When available, chemical composition similarity is also incorporated. The ranked candidates and confidence scores from all components are aggregated to produce the final phase and crystallographic symmetry predictions. While the retrieval and classifier primarily contribute to predictive performance, the IG-based module mainly supports interpretability and case-based reasoning. We refer to this aggregated multi-decision inference strategy as Hybrid Physics-Informed Decision (HyPhID) inference, which produces ranked and interpretable crystallographic predictions.
The main contributions of this work are:
(i) a unified multi-view framework combining representation learning, retrieval, and explainable reasoning for XRD phase identification;
(ii) an adaptive cosine-retrieval strategy enabling non-parametric, ranked label inference;
(iii) an explanation-guided retrieval mechanism using Integrated Gradients attribution fingerprints;
(iv) integration of chemical composition embeddings as complementary structural evidence.

\section{Synthetic XRD Dataset Generation}
To train and evaluate the proposed model under controlled and physically meaningful conditions, a large synthetic X-ray diffraction (XRD) dataset is constructed from crystallographic information files (CIFs) obtained from the Materials Project (MP)~\cite{Jain2013MaterialsProject} and the Crystallography Open Database (COD). In total, 68,958 crystal structures are collected (58,418 from MP and 10,540 from COD). The COD database primarily contains experimentally determined crystal structures, whereas the MP mainly provides density-functional-theory (DFT) relaxed structures, resulting in a dataset that combines both experimentally derived and computationally generated structures. Each CIF is treated as a distinct crystal-structure prototype from which multiple diffraction patterns are generated to emulate realistic experimental variability. For crystallographic symmetry prediction, two datasets are constructed. For space-group prediction, approximately 2000 CIF structures are sampled per space group from 22 representative space groups selected from the 230 crystallographic space groups to maintain a balanced dataset with sufficient training samples. For crystal-system prediction, approximately 7000 CIF structures are sampled for each of the seven crystal systems (triclinic, monoclinic, orthorhombic, tetragonal, trigonal, hexagonal, and cubic). For each CIF, 15 additional diffraction patterns are generated using physically motivated data augmentation. Diffraction patterns are simulated over a fixed angular range and discretized onto a uniform intensity grid. Experimental variability is simulated through physically-motivated augmentations such as peak broadening, preferred orientation, background contributions, intensity scaling, and small peak shifts, simulating realistic instrumental and sample-related conditions while preserving the underlying crystallographic phase label. Because a single composition can form multiple crystal structures (polymorphism), diffraction patterns reflect crystallographic symmetry rather than composition alone. The use of physically-motivated augmentation is particularly important for bridging the gap between simulated and experimental diffraction patterns.
In practice, experimental measurements are influenced by instrumental resolution, preferred orientation, sample preparation, strain effects, and background noise. These factors can substantially alter peak intensities and peak shapes without changing the underlying crystal symmetry. By incorporating such perturbations during data generation, the model is encouraged to learn robust structural representations rather than memorizing idealized diffraction signatures.

\section{Model Architecture and Training}
We employ a physics-informed convolutional autoencoder augmented with an auxiliary classifier for XRD phase analysis. The model jointly supports phase classification, similarity-based retrieval, and structured representation learning under realistic experimental variability. The physics-informed aspect arises from incorporating domain-specific constraints during training, including peak-aware reconstruction weighting and smoothness regularization that encourage physically plausible diffraction signals. A shared convolutional encoder extracts compact features that are disentangled into a reconstruction latent representation and a discriminative classification latent representation. The reconstruction branch enables faithful signal reconstruction, while the classification branch supports supervised learning and contrastive structuring of the latent space.
\subsection{Architecture}
The encoder consists of residual one-dimensional convolutional blocks with progressive downsampling. Early layers employ large convolutional kernels to capture broad peak envelopes and background variations, whereas deeper layers focus on fine diffraction features. Each residual block includes batch normalization, GELU activation, and an optional channel-attention mechanism that enhances informative feature channels.
Residual skip connections improve optimization stability and preserve weak but structurally meaningful reflections.
Given an input diffraction pattern $x \in \mathbb{R}^{L}$, the encoder produces a feature tensor that is flattened and mapped into two latent vectors:
\[
z_{\mathrm{rec}} \in \mathbb{R}^{d_{\mathrm{rec}}},
\qquad
z_{\mathrm{cls}} \in \mathbb{R}^{d_{\mathrm{cls}}}.
\]
Layer normalization is applied independently to both latent spaces.
The reconstruction latent $z_{\mathrm{rec}}$ encodes information required for accurate signal reconstruction, including background and noise characteristics. A fully connected expansion layer followed by a convolutional decoder with progressive upsampling reconstructs the input signal:
\[
\hat{x} \in \mathbb{R}^{L}.
\]
The classification latent $z_{\mathrm{cls}}$ captures discriminative structural features relevant for phase prediction and similarity learning. Predictions are obtained by passing $z_{\mathrm{cls}}$ through a lightweight residual multilayer perceptron with dropout
regularization. During training, an optional projection head is applied to $z_{\mathrm{cls}}$ to enable contrastive learning and improve latent space organization.

\subsection{Training Objectives}
The model is trained end-to-end using a multi-objective loss that balances reconstruction fidelity, classification accuracy, smoothness regularization, and latent space structuring.
Signal reconstruction is optimized using a peak-aware loss that assigns greater importance to diffraction peak regions:
\begin{equation}
\mathcal{L}_{\mathrm{rec}} =
\mathrm{MSE}_{\mathrm{weighted}}(x, \hat{x})
+ \lambda_{1} \| x - \hat{x} \|_{1}.
\end{equation}
Peak regions are identified using an intensity threshold relative to the maximum peak height, ensuring preservation of weak but structurally meaningful reflections. To encourage physically plausible smooth reconstructions, a gradient sparsity penalty is introduced:
\begin{equation}
\mathcal{L}_{\mathrm{smooth}} =
\left\| \nabla \hat{x} \right\|_{1},
\end{equation}
where $\nabla \hat{x}$ denotes the discrete first derivative of the reconstructed signal.
Phase prediction is optimized using cross-entropy loss:
\begin{equation}
\mathcal{L}_{\mathrm{cls}} =
- \sum_{k} y_k \log p_k .
\end{equation}
To explicitly structure the discriminative latent space, supervised contrastive learning is applied to projected embeddings:
\begin{equation}
\mathcal{L}_{\mathrm{sup}} =
- \frac{1}{N}
\sum_{i}
\frac{1}{|P(i)|}
\sum_{p \in P(i)}
\log
\frac{\exp(z_i \cdot z_p / \tau)}
{\sum_{a \neq i} \exp(z_i \cdot z_a / \tau)},
\end{equation}
where $P(i)$ denotes samples sharing the same phase label as sample $i$, and $\tau$ is a temperature parameter. This objective encourages samples belonging to the same phase to cluster in latent space, improving similarity-based retrieval and robustness to structural ambiguity.
The overall loss function is defined as
\begin{equation}
\mathcal{L} =
\lambda_{\mathrm{rec}} \mathcal{L}_{\mathrm{rec}} +
\lambda_{\mathrm{cls}} \mathcal{L}_{\mathrm{cls}} +
\lambda_{\mathrm{smooth}} \mathcal{L}_{\mathrm{smooth}} +
\lambda_{\mathrm{sup}} \mathcal{L}_{\mathrm{sup}}.
\end{equation}
The weighting coefficients $\lambda_{\mathrm{rec}}$, $\lambda_{\mathrm{cls}}$, $\lambda_{\mathrm{smooth}}$, and $\lambda_{\mathrm{sup}}$ correspond to the reconstruction loss, classification loss, smoothness regularization term, and supervised contrastive loss, respectively. The weights are selected empirically using validation experiments to balance reconstruction and classification performance.
\section{Integrated Gradients Database for Explainable Case-Based Prediction}
\label{sec:ig_database}
Reliable XRD phase identification requires interpretability aligned with physics-based reasoning. We therefore integrate Integrated Gradients (IG) explanations into a retrieval-based prediction framework. For a diffraction pattern $x \in \mathbb{R}^{L}$ and model $F(x)$, IG assigns feature importance as
\begin{equation}
\phi_i(x) =
(x_i - x'_i)
\int_{0}^{1}
\frac{\partial F(x' + \alpha (x - x'))}{\partial x_i}
\, d\alpha,
\end{equation}
where $x'$ is a zero-intensity baseline and the integral is approximated numerically. For each training sample, we compute
$\boldsymbol{\phi} = \mathrm{IG}(x, y_{\text{true}})$
using the ground-truth class as attribution target. All attribution vectors are stored together with their labels and dataset indices.
To enable efficient similarity search, attributions are mean-centered and $\ell_2$-normalized:
\begin{equation}
\tilde{\boldsymbol{\phi}} =
\frac{\boldsymbol{\phi} - \mathrm{mean}(\boldsymbol{\phi})}
{\|\boldsymbol{\phi} - \mathrm{mean}(\boldsymbol{\phi})\|_2}.
\end{equation}
During inference, IG explanations are computed for a query pattern when additional interpretability is required (e.g., high uncertainty or model-retrieval disagreement). Explanation similarity is measured via cosine similarity,
\begin{equation}
S_{\mathrm{IG}}(q,i) =
\tilde{\boldsymbol{\phi}}(x_q)^\top
\tilde{\boldsymbol{\phi}}(x_i),
\end{equation}
and the top-$k$ neighbors contribute class evidence through
similarity-weighted voting:
\begin{equation}
V_c =
\sum_{i \in \mathcal{N}_k}
\max(0, S_{\mathrm{IG}}(q,i))
\, \mathbf{1}[y_i = c].
\end{equation}
Class probabilities are obtained via temperature-scaled softmax,
\begin{equation}
P_c =
\frac{\exp(V_c / \tau)}
{\sum_j \exp(V_j / \tau)}.
\end{equation}
This procedure yields an explanation-driven prediction based on similarity of attribution patterns, enabling case-based reasoning in attribution space.
\section{Adaptive Cosine Retrieval for Latent-Space Classification}
We perform non-parametric classification directly in the latent space of a trained autoencoder. After training, the encoder is frozen and used as a deterministic feature extractor. All training samples are embedded once to form a latent reference database 
$\mathcal{D}_{\text{train}}=\{(\mathbf{z}_i,y_i)\}_{i=1}^{N}$.
Given a query sample $x_q$, its latent representation $\mathbf{z}_q=f(x_q)$ is $\ell_2$-normalized together with all stored latents. Classification is then performed via cosine-similarity retrieval. Instead of using a fixed number of neighbors, we evaluate multiple neighborhood sizes and adaptively select the most informative one: the neighborhood that yields the most decisive prediction, measured by the highest top-1 class probability.
For each candidate neighborhood size, class evidence is accumulated by summing similarity contributions of neighbors belonging to the same class. Negative similarities may optionally be ignored to prevent contradictory evidence. The aggregated class scores are converted into probabilities using a temperature-scaled softmax, where the temperature parameter $T$ controls the sharpness of the probability distribution (lower values produce sharper class probabilities). The neighborhood size producing the most decisive prediction (highest top-1 probability) is selected, and the corresponding label is returned together with the top-$n$ candidate classes for interpretability. Algorithm~\ref{alg:adaptive_cosine_retrieval} summarizes the retrieval procedure. In the pseudocode, ``$\leftarrow$'' denotes variable assignment, and ``None'' indicates that no prediction has been selected yet.
\begin{algorithm}[t]
\small
\caption{Adaptive Cosine Retrieval with Decisive-$k$ Selection}
\label{alg:adaptive_cosine_retrieval}
\KwIn{
Query sample $x_q$; encoder $f$; latent database
$\mathcal{D}_{\text{train}}$; candidate sizes $\mathcal{K}$;
similarity threshold $\tau_s$; temperature $T$; top-$n$ size $n$}
\KwOut{Predicted label $\hat{y}$ and top-$n$ candidates}
Compute latent $\mathbf{z}_q=f(x_q)$ and normalize all latents\;
Compute cosine similarities $s_i$ and sort in descending order\;
Initialize best prediction $\leftarrow$ None\;
\ForEach{$k \in \mathcal{K}$}{
  Select top-$k$ neighbors $\mathcal{N}_k$\;
  Compute average similarity
  $\bar{s}_k=\frac{1}{k}\sum_{i\in\mathcal{N}_k}s_i$\;
   \If{$\bar{s}_k < \tau_s$}{\textbf{break}}
  
Aggregate class scores
  $S_c(k)=\sum_{i\in\mathcal{N}_k,\;y_i=c}\tilde{s}_i$\;
  
  Compute probabilities
  $p_c(k)=\text{softmax}(S_c(k)/T)$\;
  
  Update best result if $\max_c p_c(k)$ increases\;
}
Return label with highest probability and top-$n$ candidates\;
\end{algorithm}

\section{Embedding-Based Chemical Retrieval}
\label{sec:retrieval}
We investigate whether learned embeddings of chemical compositions capture meaningful chemical similarity. Given a query composition, the goal is to retrieve chemically similar compositions from a reference database based on similarity in the learned embedding space. Since each composition corresponds to a material with a known crystal structure (defined by its crystal system and space group), composition-based retrieval can provide complementary structural evidence for phase identification. Chemical composition embeddings are generated using the Qwen2.5-1.5B-Instruct model~\cite{yang2024qwen2}. To study different representation granularities, we compare two learned encoding strategies with a physically motivated baseline based on Magpie descriptors. In the element-level encoder, a composition is represented as a weighted sum of element embeddings scaled by their stoichiometric fractions. The composition-level encoder embeds the full chemical formula as a textual sequence using the Qwen model. Embeddings are extracted from the final transformer layer and mean-pooled, yielding fixed-dimensional vectors ($d=2048$). As a physically motivated baseline, we use Magpie descriptors~\cite{ward2016magpie}, which capture statistical properties of constituent elements (e.g., atomic number and electronegativity). Features are imputed, standardized, and represented as 145-dimensional vectors. All formulas are converted to a canonical representation and $\ell_2$-normalized prior to similarity computation.
Nearest neighbors are retrieved using cosine similarity between normalized representations. For each query composition, the top-$k$ most similar training compositions are selected. Retrieval performance is evaluated using \emph{chemical accuracy@k}, defined as the fraction of retrieved compositions sharing at least half of their constituent elements with the query. We additionally report \emph{retrieval overlap@k}, which quantifies agreement between the element-level and composition-level encoders. Both metrics are applied consistently across learned embeddings and the Magpie baseline.

\section{Results and Discussion}
This section evaluates the proposed framework from three complementary perspectives. First, we analyze model decision behavior using Integrated Gradients to assess whether the learned representations rely on physically meaningful diffraction features. Second, we evaluate the effectiveness of composition-based retrieval using learned chemical embeddings. Finally, we report classification performance for crystal system and space group prediction and discuss how the different inference components contribute
to the final decision.

\subsection{Integrated Gradients Analysis of Model Decisions}
\begin{figure}[t]
    \centering
    \begin{subfigure}[t]{0.48\linewidth}
        \centering
        \includegraphics[width=1\linewidth]
        {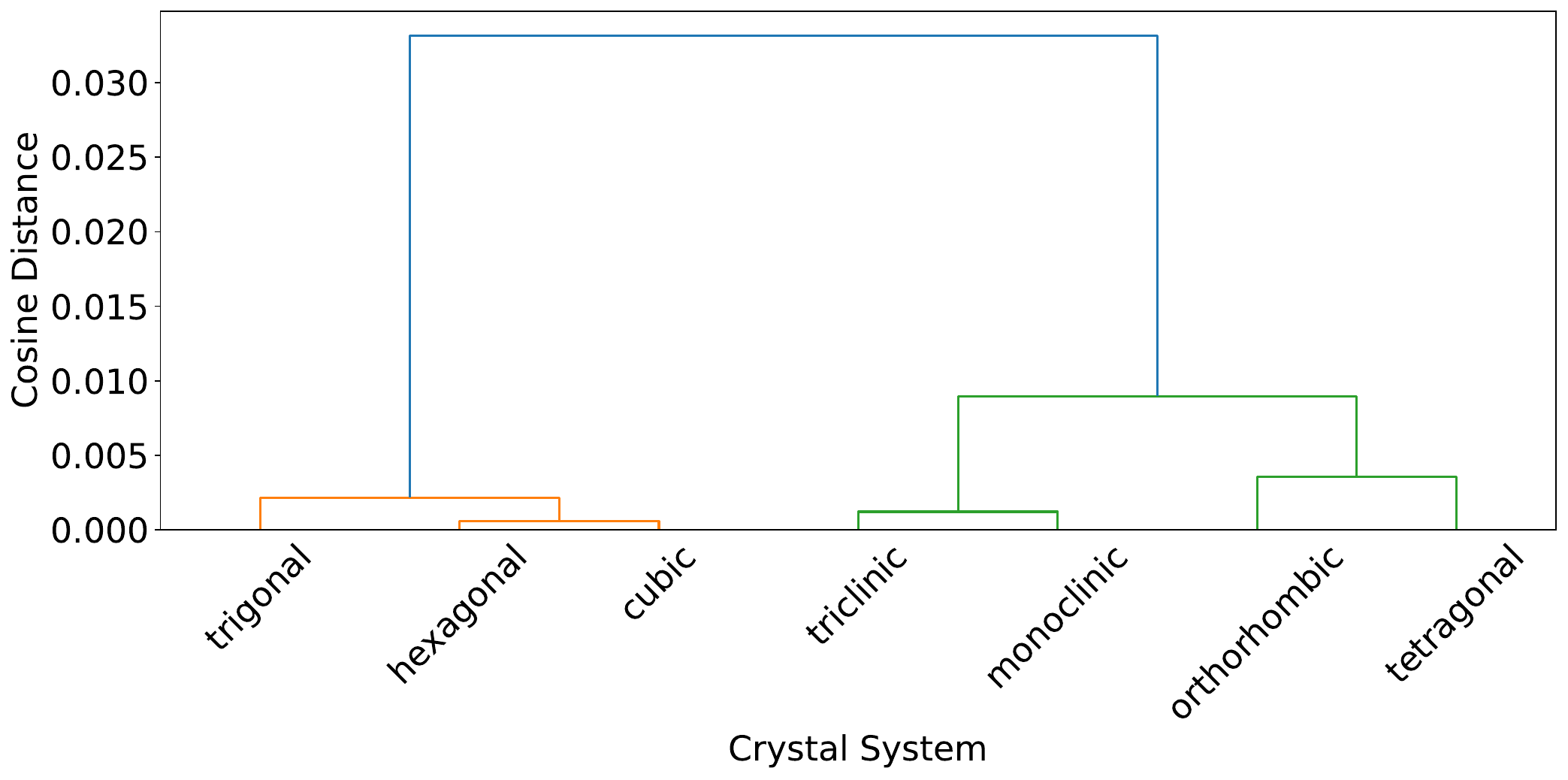}
        \caption{Hierarchical clustering of crystal systems based on cosine distances between normalized attribution fingerprints.}
        \label{fig:ig_clustering_crystalsystems}
    \end{subfigure}
    \hfill
    \begin{subfigure}[t]{0.48\linewidth}
        \centering
        \includegraphics[width=1\linewidth]
        {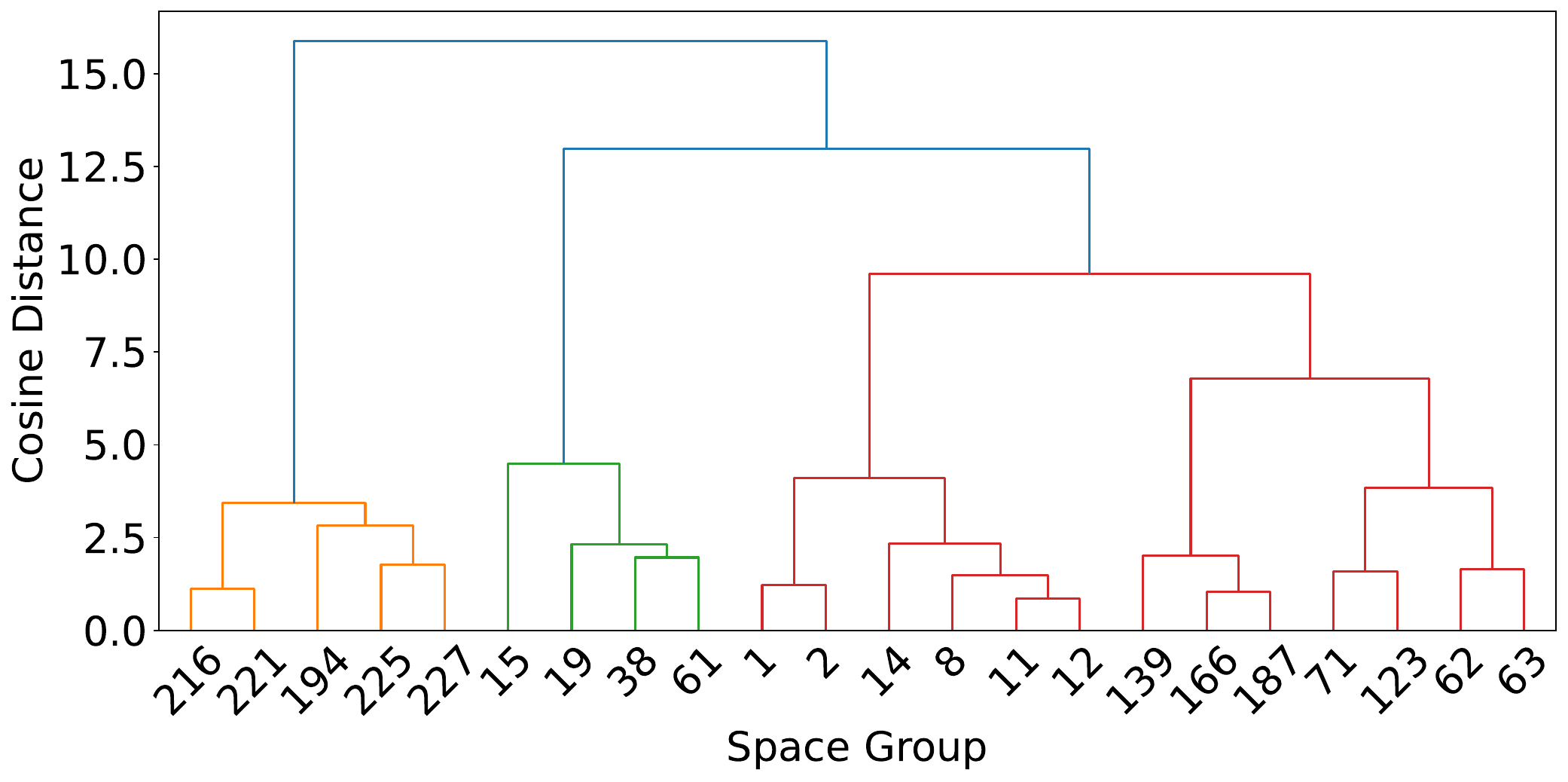}
        \caption{Hierarchical clustering of space groups based on cosine distances between normalized Integrated Gradients attribution fingerprints.}
        \label{fig:ig_clustering_spacegroups}
    \end{subfigure}
   \caption{Hierarchical clustering of crystallographic classes derived from cosine distances between normalized Integrated Gradients attribution fingerprints: (a) crystal systems and (b) space groups.}
    \label{fig:ig_clustering_combined}
\end{figure}

To explain the classification behavior, we analyze model decisions using Integrated Gradients (IG). IG attributions are aggregated across correctly classified samples to construct \emph{class-level attribution fingerprints}. For each diffraction pattern, absolute IG values are computed along the $2\theta$ domain and averaged within each class to produce normalized importance profiles indicating influential diffraction regions. Relationships between crystallographic classes are quantified by computing cosine distances between attribution fingerprints and applying hierarchical clustering. The resulting dendrograms (Figure~\ref{fig:ig_clustering_combined}) show space-group clustering on the left and crystal-system clustering on the right. The clusters follow crystallographic similarity rather than training labels, suggesting that explanation patterns capture underlying structural relationships. Higher-symmetry systems tend to group together due to reliance on dominant diffraction peaks, whereas lower-symmetry systems form broader clusters reflecting more distributed diffraction signatures.
Overall, IG analysis indicates that model predictions rely on physically meaningful diffraction features and provides interpretable evidence linking learned representations with crystallographic structure. The observed attribution patterns are consistent with crystallographic expectations, where a limited number of characteristic reflections often carry substantial information about lattice symmetry. This suggests that the model is learning physically meaningful decision rules rather than relying on dataset-specific artifacts. Furthermore, the attribution fingerprints provide an alternative representation of diffraction patterns that can be compared independently from the latent
representations learned by the autoencoder. This motivates the use of
attribution-based retrieval as an additional source of evidence during
inference.

\subsection{Results of Embedding-Based Chemical Retrieval}
\begin{figure}[t]
    \centering
    \begin{subfigure}[t]{0.48\linewidth}
        \centering
        \includegraphics[width=1\linewidth]{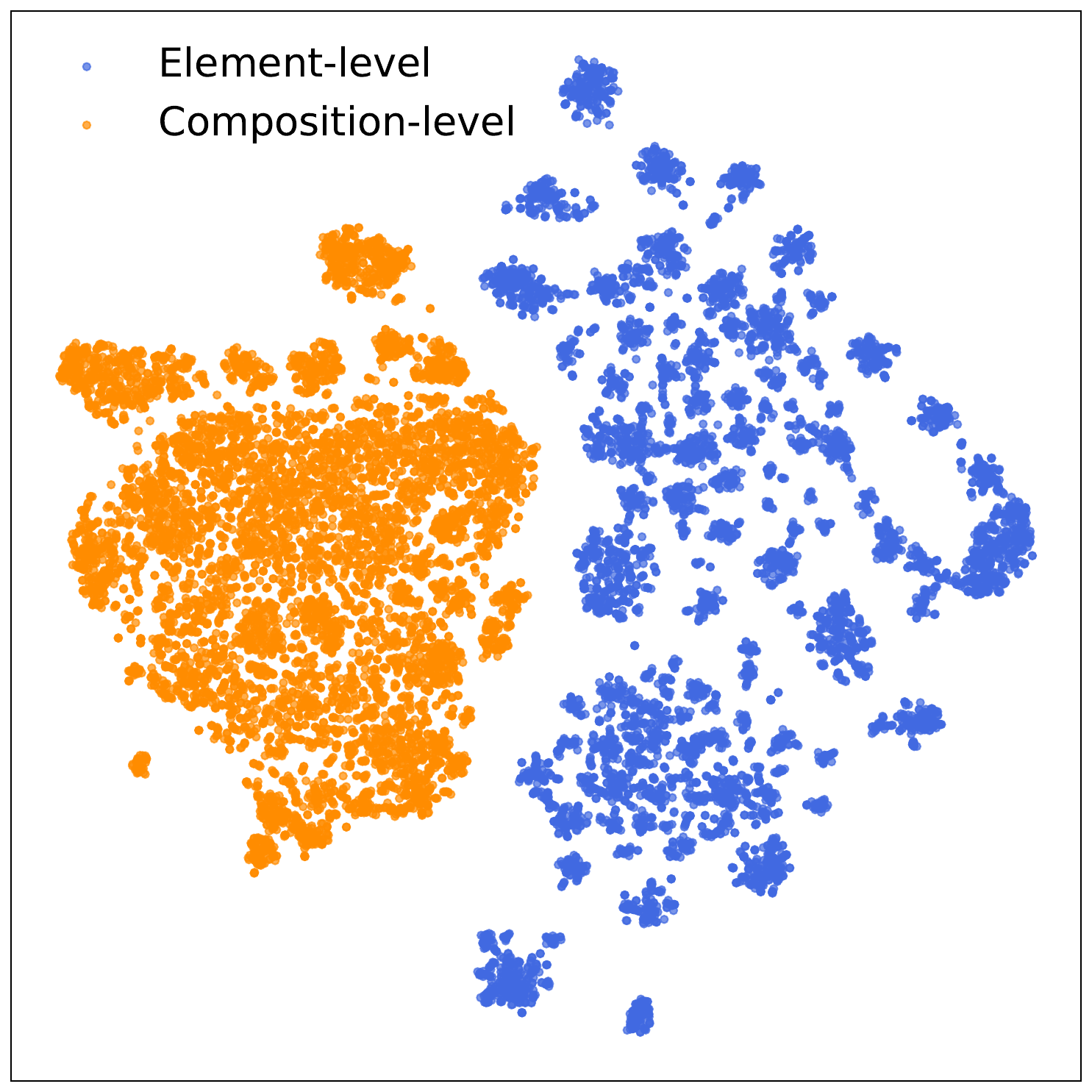}
        \caption{Qwen element-level vs composition-level embeddings.}
        \label{fig:tsne_qwen}
    \end{subfigure}
    \hfill
    \begin{subfigure}[t]{0.48\linewidth}
        \centering
        \includegraphics[width=1\linewidth]{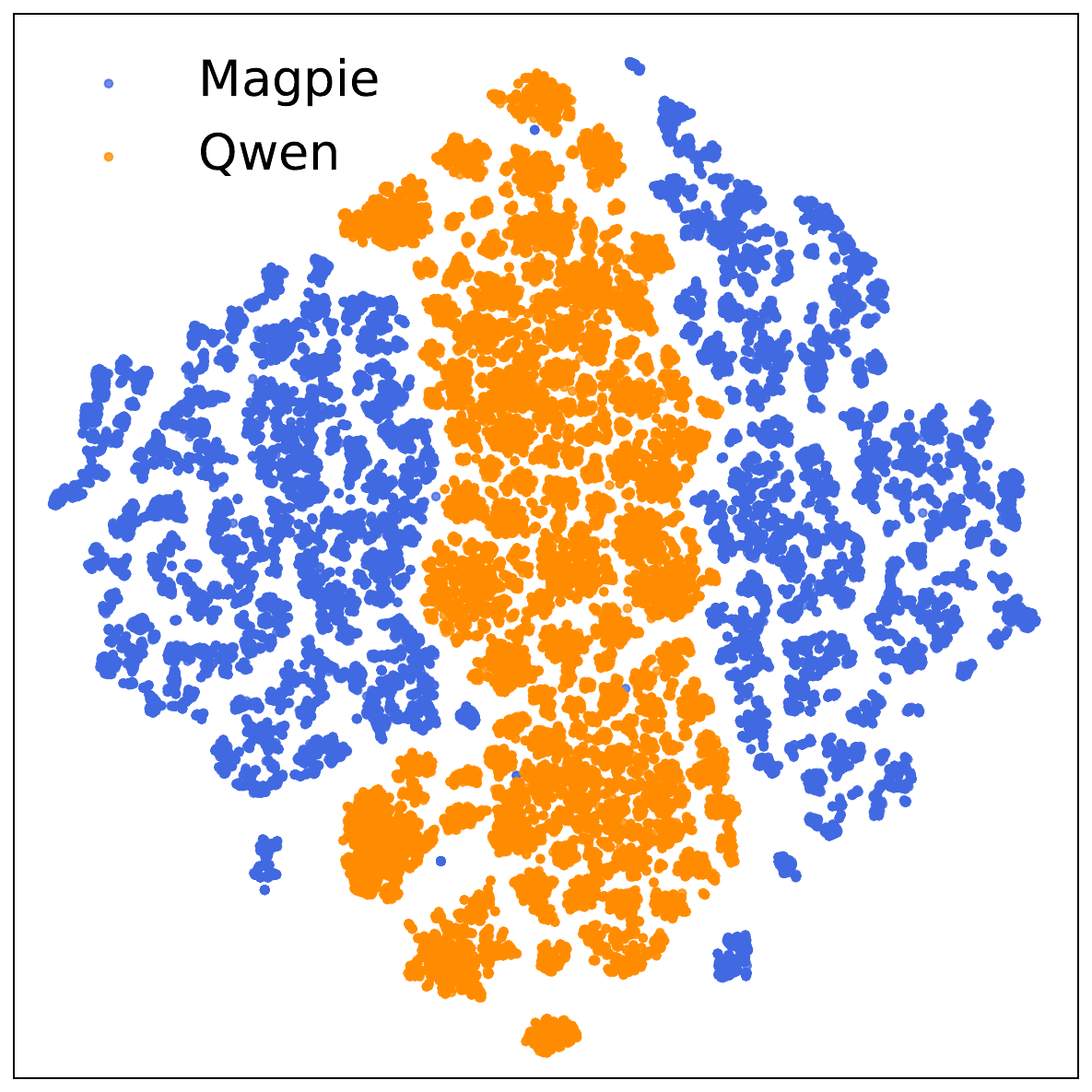}
        \caption{Magpie descriptors vs Qwen element-level embeddings after alignment.}
        \label{fig:tsne_magpie_qwen}
    \end{subfigure}
    \caption{t-SNE visualization of chemical embedding spaces.
(a) Element-level vs. composition-level Qwen embeddings.
(b) Magpie descriptors compared with Qwen element-level embeddings
after alignment to a shared representation space.
}
    \label{fig:tsne_magpie_qwen_all}
\end{figure}

We evaluate compositional retrieval using the metrics defined in
Section~\ref{sec:retrieval}. Table~\ref{tab:qwen_vs_qwen} compares element-level and composition-level Qwen embeddings. Element-level embeddings achieve higher chemical accuracy@5 and substantially lower variance, indicating more stable retrieval of chemically similar compositions. The low retrieval overlap and weak embedding correlation further suggest that the two encoders capture different notions of chemical similarity. We find the element-level encoder outperforms the composition-level encoder, so we compare the element-level encoder with the conventional Magpie descriptors~\cite{ward2016magpie} and report the results in Table~\ref{tab:magpie_vs_qwen}.
Element-level Qwen embeddings outperform Magpie features in both accuracy and stability, demonstrating the advantage of learned representations over hand-engineered descriptors.
Figure~\ref{fig:tsne_magpie_qwen_all} visualizes the chemical embedding spaces using t-SNE, including Qwen element-level embeddings, Qwen composition-level embeddings, and Magpie descriptor representations. Element-level embeddings form compact clusters consistent with element-based similarity, whereas composition-level embeddings exhibit a more diffuse organization. To enable joint visualization, Magpie and Qwen embeddings are first aligned by composition, meaning that identical compositions are matched across the different representations. Each representation is then independently projected to a lower-dimensional space using principal component analysis (PCA) to reduce dimensionality while preserving the variance structure of each representation. The PCA-reduced representations are then concatenated and embedded using t-SNE, allowing both learned and hand-crafted representations to be visualized within a shared map. The resulting visualization reveals clearly distinct structural layouts between Magpie descriptors and the learned Qwen embeddings. Similar differences between learned and descriptor-based chemical representations have been discussed in recent work on embedding spaces and linear mixing behavior~\cite{zhang2025electrocatalyst}.
Based on the quantitative retrieval results in Table~\ref{tab:magpie_vs_qwen}, we adopt element-level Qwen embeddings as the compositional retrieval module in our phase identification framework. During inference, a query composition retrieves the top-$k$ chemically similar references using cosine similarity, and the associated crystallographic metadata (crystal system and space group) is used as complementary structural evidence to support diffraction-based prediction.
\begin{table}[t]
\centering
\setlength{\tabcolsep}{3pt} 
\caption{Chemical retrieval performance of Qwen embeddings ($k=5$).}
\label{tab:qwen_vs_qwen}
\begin{tabular}{lcc}
\hline
\textbf{Metric} & \textbf{Element-level} & \textbf{Composition-level} \\
\hline
Chemical accuracy@5 (mean) & 0.985 & 0.887 \\
Chemical accuracy@5 (std)  & 0.093 & 0.227 \\
Retrieval overlap@5        & \multicolumn{2}{c}{0.167} \\
Embedding space correlation & \multicolumn{2}{c}{0.241} \\
\hline
\end{tabular}
\end{table}

\begin{table}[t]
\centering
\renewcommand{\arraystretch}{1.15}
\setlength{\tabcolsep}{3pt}
\caption{Retrieval performance of Magpie vs. Qwen embeddings ($k=5$).}
\label{tab:magpie_vs_qwen}
\begin{tabular}{lcc}
\hline
\textbf{Embedding method} & \textbf{Chemical accuracy@5 (mean)} & \textbf{Std} \\
\hline
Magpie descriptors        & 0.843          & 0.244 \\
Qwen (element-level)      & \textbf{0.982} & 0.093 \\
\hline
\end{tabular}
\end{table}

\subsection{Classification Performance}
\begin{figure}[t]
\centering
\begin{subfigure}[t]{0.48\linewidth}
    \centering
    \includegraphics[width=1\linewidth]{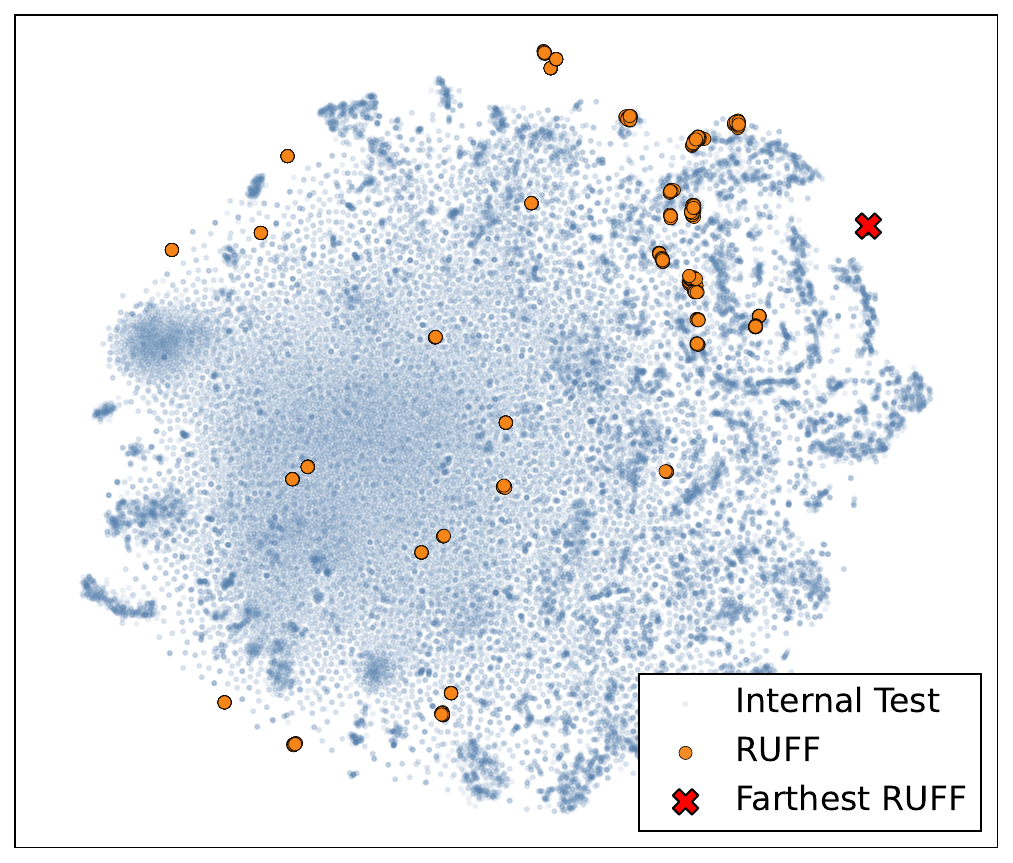}
    \caption{Latent-space visualization of training samples, internal test samples, and external RRUFF samples. Red crosses indicate the most distant samples used for robustness evaluation.}
    \label{fig:latent_space}
\end{subfigure}
\hfill
\begin{subfigure}[t]{0.48\linewidth}
    \centering
    \includegraphics[width=1\linewidth]{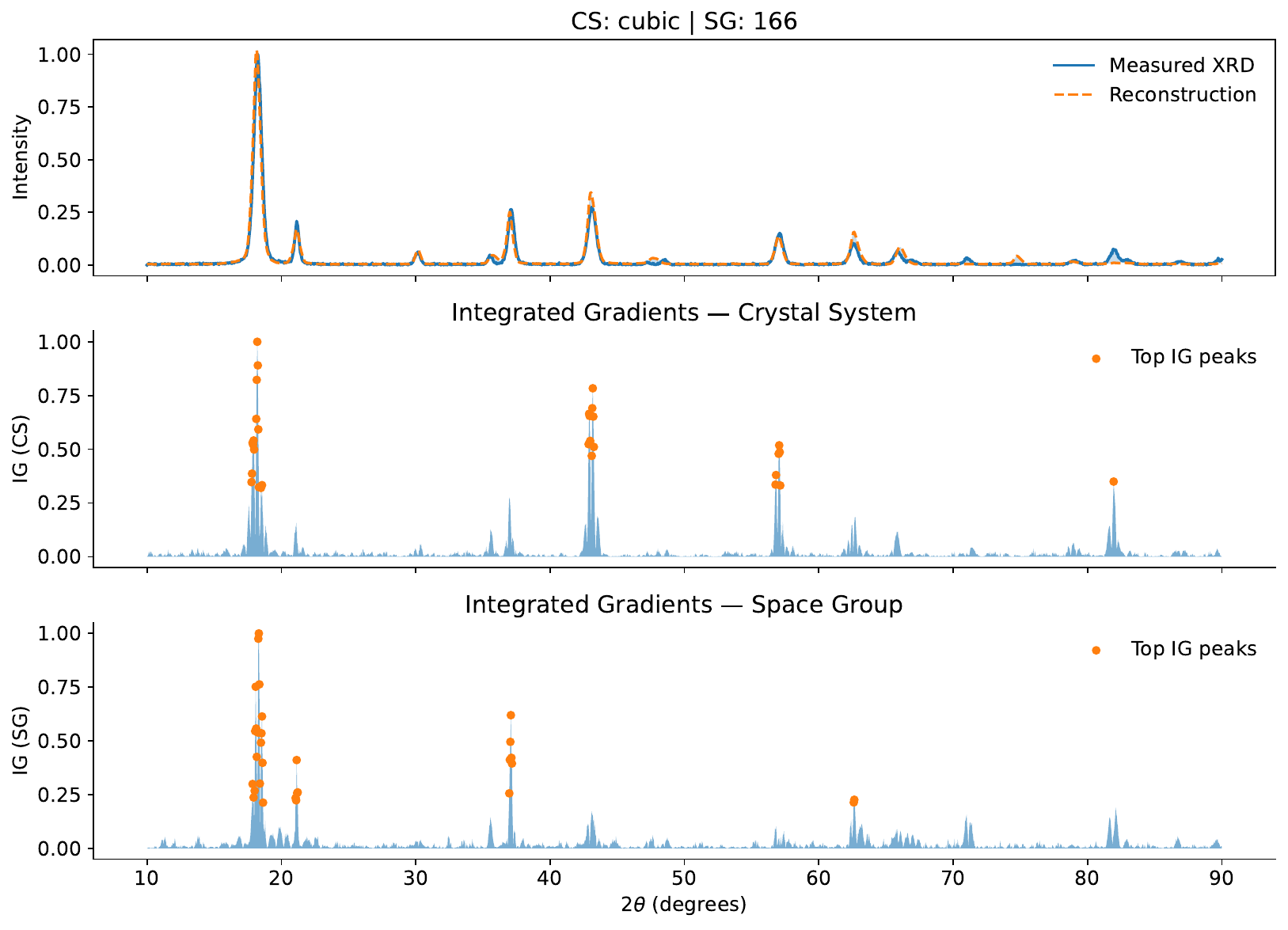}
    \caption{Representative prediction example showing the measured XRD pattern, reconstruction, and Integrated Gradients (IG) attribution maps for crystal system and space group prediction.}
    \label{fig:Rec_IG}
\end{subfigure}

\caption{Latent-space visualization and example HyPhID inference.
(a) Distribution of training, internal test, and external RRUFF samples. (b) Example prediction showing the measured pattern,
reconstruction, and IG-based peak attribution.}
\label{fig:latent_and_inference}
\end{figure}

We evaluate our framework on two tasks: crystal system classification and space group classification, using a 10\% randomly selected holdout set for both tasks. To avoid data leakage, the split is performed at the CIF level, ensuring that all augmented diffraction patterns derived from the same structure remain within the same partition. The model achieves 98.85\% accuracy for crystal system prediction. Space group prediction is more challenging due to the finer structural distinctions among the 22 symmetry classes; nevertheless, the framework achieves 95.82\% accuracy on the same holdout set. Both tasks are evaluated using the same encoder architecture and training procedure. The only difference between the two models is the final classification layer, which is adjusted to predict either the 7 crystal systems or the 22 space groups.
To further assess robustness, we evaluate the framework on an external dataset of 50 CIF structures from the experimental RRUFF database~\cite{downs2006rruff}. These compositions are unseen during training and internal testing. For each CIF, 15 augmented diffraction patterns are generated, resulting in 750 evaluation samples. Unlike the COD/MP training dataset, which includes both experimentally derived and DFT-relaxed structures, the RRUFF dataset contains experimentally reported, non-relaxed structures, making it a more challenging test scenario.
As expected, performance on this external dataset is lower than on the
internal holdout set due to the domain shift between simulated and
experimentally measured diffraction patterns. On the internal holdout
set, HyPhID achieves 98.85\% accuracy for crystal system prediction and
95.82\% accuracy for space group prediction, while on the experimental
RRUFF dataset the accuracy decreases to 81.07\% and 56.13\%, respectively.
The substantial performance gap between the internal and external evaluations highlights the challenge of generalization in diffraction analysis. While synthetic datasets enable large-scale supervised
training, experimentally measured diffraction patterns often exhibit sources of variability that are difficult to model completely. These include instrumental effects, preferred orientation, peak broadening,
background artifacts, and sample impurities. As a result, performance on external datasets provides a more realistic estimate of practical
applicability.
The larger performance reduction observed for space-group prediction is expected because space groups represent much finer crystallographic distinctions than crystal systems. Many space groups share similar peak positions and differ only through subtle intensity variations or weak reflections that may be difficult to distinguish in noisy experimental
measurements.
In comparison, Vecsei et al.~\cite{vecsei2019neural} reported 85\% crystal-system accuracy and 76\% space-group accuracy on simulated data, with performance dropping to 56\% and 42\% on the experimental RRUFF dataset. Table~\ref{tab:ablation_methods} summarizes the performance of the individual decision components together with the final HyPhID inference results on the RRUFF dataset. Interestingly, latent retrieval slightly outperforms the standalone
classifier on the RRUFF dataset for both crystal-system and space-group prediction. This observation suggests that similarity-based reasoning is less sensitive to domain shift than purely parametric classification. Rather than extrapolating from learned decision boundaries, retrieval directly compares a query pattern against structurally related examples, allowing physically similar reference structures to contribute to the final decision.
In contrast, the IG and composition modules exhibit lower standalone accuracies. This behavior is expected because these components are not
designed to function as independent classifiers. Instead, they provide complementary evidence that can support or refine predictions produced by the classifier and retrieval modules. The final HyPhID prediction benefits from aggregating these diverse sources of information.
In this evaluation, we report the predictions of all HyPhID decision modules on the external test set, since the multi-decision inference strategy improves robustness by combining complementary reasoning sources. Because the retrieval-based modules are based on similarity to reference patterns, their standalone accuracies may vary compared to the classifier performance observed on the synthetic holdout set.
The final HyPhID prediction aggregates normalized scores from all decision components. Let $p_{\text{model}}(c)$, $p_{\text{retrieval}}(c)$, $p_{\text{IG}}(c)$, and $p_{\text{comp}}(c)$ denote the class probabilities produced by the classifier, latent retrieval, IG-based similarity, and composition retrieval modules, respectively. The final class score is computed as
\[
S(c) = w_1 p_{\text{model}}(c) +
       w_2 p_{\text{retrieval}}(c) +
       w_3 p_{\text{IG}}(c) +
       w_4 p_{\text{comp}}(c),
\]
where the weights satisfy $w_1 + w_2 + w_3 + w_4 = 1$.
The weights are selected empirically using validation test sets, and the final prediction is obtained as $\arg\max_c S(c)$.
Latent-space retrieval often produces reliable predictions and can correctly classify samples even when the classifier fails, highlighting the value of similarity-based reasoning. The IG module mainly supports interpretability, while composition-based retrieval provides complementary structural evidence despite polymorphism effects. Overall, HyPhID improves crystal system prediction by combining complementary evidence sources.
To analyze robustness under distribution shift, we examine the most distant RRUFF samples in latent space. Figure~\ref{fig:latent_and_inference}(a) visualizes the distribution of training samples, internal test samples, and external RRUFF structures in the learned latent representation, where the most distant samples are marked with red crosses. An example inference case for one of these distant samples is illustrated in Figure~\ref{fig:latent_and_inference}(b), showing the reconstructed diffraction pattern and IG attribution maps highlighting the most influential peaks. Finally, Figure~\ref{fig:representative_inference} presents a detailed inference output example summarizing the predictions produced by the classifier, latent retrieval, IG-based similarity, and composition-based reasoning, together with the top-$k$ candidate classes returned by each component. This example illustrates disagreement between inference components. While the correct space group is SG~225, the retrieval module favors SG~166. However, the correct class (SG~225) still appears among the top retrieved candidates. Importantly, our framework is trained in a composition-agnostic manner and does not require explicit composition-specific supervision during training. This contrasts with approaches such as~\cite{szymanski2021probabilistic}, which incorporate composition constraints or probabilistic modeling of phase mixtures. Instead, HyPhID enables flexible inference by combining multiple evidence sources and returning a ranked list of plausible crystallographic classes, which can assist users in narrowing the search space for possible structures even when the top-$1$ prediction is uncertain.

\begin{table}[!t]
\centering
\footnotesize
\setlength{\tabcolsep}{3pt}
\renewcommand{\arraystretch}{0.9}
\caption{Accuracy of individual decision components and the final HyPhID inference for crystal system (CS) and space group (SG) classification.}
\label{tab:ablation_methods}
\begin{tabular}{lcc}
\toprule
\textbf{Method} & \textbf{CS Acc. (\%)} & \textbf{SG Acc. (\%)} \\
\midrule
Classifier Model & 76.13\% & 50.80\% \\
Latent Retrieval & 78.13\% & 56.67\% \\
IG Similarity    & 52.93\% & 16.80\% \\
Composition      & 32.00\% & 22.00\% \\
\midrule
\textbf{HyPhID (Final)} & \textbf{81.07\%} & \textbf{56.13\%} \\
\bottomrule
\end{tabular}
\end{table}

\begin{figure}[t]
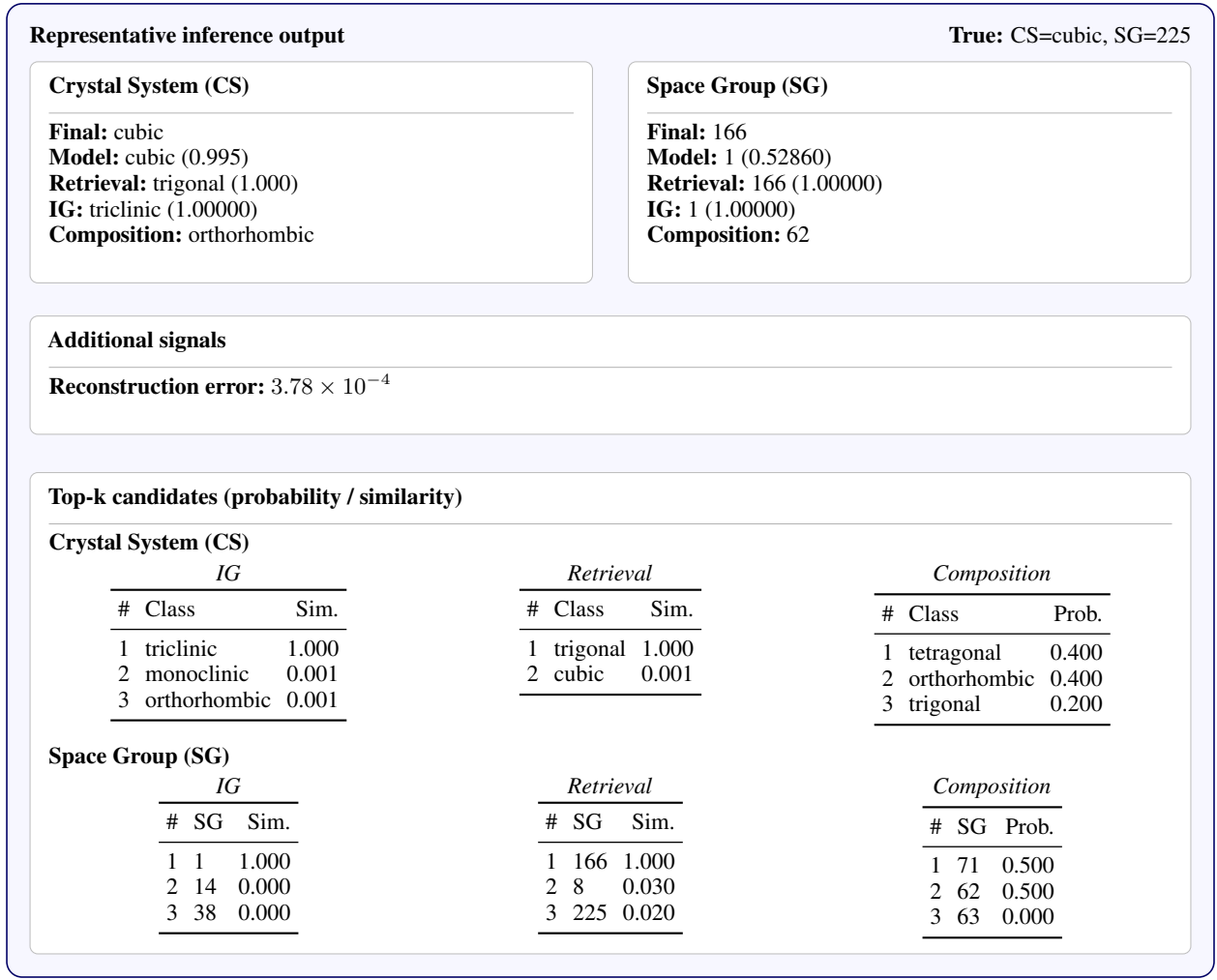

\centering

\begin{inferbox}
\small
\textbf{Representative inference output}\hfill
\textbf{True:} CS=cubic, SG=225

\vspace{0.5em}

\begin{minipage}[t]{0.485\linewidth}
\begin{inferblock}
\textbf{Crystal System (CS)}\\[-2pt]
{\color{black!30}\rule{\linewidth}{0.3pt}}
\kv{Final}{cubic}
\kv{Model}{cubic (0.995)}
\kv{Retrieval}{trigonal (1.000)}
\kv{IG}{triclinic (1.00000)}
\kv{Composition}{orthorhombic}
\end{inferblock}
\end{minipage}
\hfill
\begin{minipage}[t]{0.485\linewidth}
\begin{inferblock}
\textbf{Space Group (SG)}\\[-2pt]
{\color{black!30}\rule{\linewidth}{0.3pt}}
\kv{Final}{166}
\kv{Model}{1 (0.52860)}
\kv{Retrieval}{166 (1.00000)}
\kv{IG}{1 (1.00000)}
\kv{Composition}{62}
\end{inferblock}
\end{minipage}

\vspace{0.5em}

\begin{inferblock}
\textbf{Additional signals}\\[-2pt]
{\color{black!30}\rule{\linewidth}{0.3pt}}
\kv{Reconstruction error}{$3.78\times10^{-4}$}
\end{inferblock}

\vspace{0.5em}

\begin{inferblock}
\textbf{Top-k candidates (probability / similarity)}\\[-2pt]
{\color{black!30}\rule{\linewidth}{0.3pt}}

\footnotesize
\setlength{\tabcolsep}{3pt}

\textbf{Crystal System (CS)}\par\vspace{2pt}

\noindent
\begin{minipage}[t]{0.32\linewidth}\centering
\textit{IG}

\vspace{2pt}
\begin{tabular}{r l r}
\toprule
\# & Class & Sim.\\
\midrule
1 & triclinic & 1.000\\
2 & monoclinic & 0.001\\
3 & orthorhombic & 0.001\\
\bottomrule
\end{tabular}
\end{minipage}
\hfill
\begin{minipage}[t]{0.32\linewidth}\centering
\textit{Retrieval}

\vspace{2pt}
\begin{tabular}{r l r}
\toprule
\# & Class & Sim.\\
\midrule
1 & trigonal & 1.000\\
2 & cubic & 0.001\\
\bottomrule
\end{tabular}
\end{minipage}
\hfill
\begin{minipage}[t]{0.32\linewidth}\centering
\textit{Composition}

\vspace{2pt}
\begin{tabular}{r l r}
\toprule
\# & Class & Prob.\\
\midrule
1 & tetragonal & 0.400\\
2 & orthorhombic & 0.400\\
3 & trigonal & 0.200\\
\bottomrule
\end{tabular}
\end{minipage}

\vspace{0.8em}

\textbf{Space Group (SG)}\par\vspace{2pt}

\noindent
\begin{minipage}[t]{0.32\linewidth}\centering
\textit{IG}

\vspace{2pt}
\begin{tabular}{r l r}
\toprule
\# & SG & Sim.\\
\midrule
1 & 1 & 1.000\\
2 & 14 & 0.000\\
3 & 38 & 0.000\\
\bottomrule
\end{tabular}
\end{minipage}
\hfill
\begin{minipage}[t]{0.32\linewidth}\centering
\textit{Retrieval}

\vspace{2pt}
\begin{tabular}{r l r}
\toprule
\# & SG & Sim.\\
\midrule
1 & 166 & 1.000\\
2 & 8 & 0.030\\
3 & 225 & 0.020\\
\bottomrule
\end{tabular}
\end{minipage}
\hfill
\begin{minipage}[t]{0.32\linewidth}\centering
\textit{Composition}

\vspace{2pt}
\begin{tabular}{r l r}
\toprule
\# & SG & Prob.\\
\midrule
1 & 71 & 0.500\\
2 & 62 & 0.500\\
3 & 63 & 0.000\\
\bottomrule
\end{tabular}
\end{minipage}

\end{inferblock}
\end{inferbox}

\caption{
Representative inference example showing disagreement between inference
sources. The model predicts the correct crystal system (cubic), while
retrieval and IG signals suggest alternative symmetries. For space-group
prediction, retrieval selects SG~166 although the ground truth is SG~225.
}
\label{fig:representative_inference}

\end{figure}

\section{Conclusion}
We present an explainable and retrieval-enhanced framework for automated phase identification from X-ray diffraction (XRD) data that combines representation learning, similarity-based retrieval, and interpretable decision support. A convolutional autoencoder learns a discriminative latent representation of diffraction patterns through joint reconstruction and classification objectives. Building on this representation, we introduce an adaptive cosine-retrieval strategy that performs non-parametric inference through class-wise similarity aggregation, enabling robust prediction beyond purely parametric classifiers. To improve transparency and scientific validity, we incorporate Integrated Gradients into a peak-level explanation database and develop explanation-guided retrieval, allowing predictions to be supported by explicit diffraction evidence aligned with expert reasoning.
In addition, compositional retrieval using large language model embeddings provides chemically-grounded prior information that complements diffractogram-based inference. Evaluation on experimentally measured XRD patterns demonstrates strong performance for both crystal system and space group prediction while yielding interpretable decision mechanisms consistent with crystallographic principles. Although our work focuses on 22 space groups, the framework can be extended to all 230 crystallographic space groups if sufficient data are available.
Overall, our framework bridges data-driven learning with domain-aware reasoning, enabling reliable and interpretable materials characterization. Future work will focus on predicting previously unseen compounds, incorporating uncertainty-aware retrieval, mixed-phase signals, and integration with materials knowledge graphs for autonomous discovery workflows. In addition, we aim to extend the framework into a publicly accessible reference platform that enables automated and explainable phase retrieval, allowing researchers to upload diffraction patterns and obtain ranked predictions supported by physically interpretable evidence.

The implementation accompanying this study, which specifies all architectural design choices, training hyperparameters, data augmentation procedures, and inference settings, is available at:
\url{https://github.com/lab-mids/xrd_classification}
\section{Acknowledgements}
All authors gratefully acknowledge funding by the Deutsche Forschungsgemeinschaft (DFG, German Research Foundation) for CRC1625–A05, INF, project number 506711657.

\bibliographystyle{unsrt}  
\bibliography{references}

@article{wang2019rapid,
  title={Rapid identification of X-ray diffraction spectra based on very limited data by interpretable convolutional neural networks},
  author={Wang, H. and Xie, Y. and Li, D. and Deng, H. and Zhao, Y. and Xin, M. and Lin, J.},
  journal={arXiv preprint arXiv:1912.07750},
  year={2019}
}

@article{szymanski2021probabilistic,
  title={Probabilistic deep learning approach to automate the interpretation of multi-phase diffraction spectra},
  author={Szymanski, N. J. and Bartel, C. J. and Zeng, Y. and Tu, Q. and Ceder, G.},
  journal={Chemistry of Materials},
  volume={33},
  number={11},
  pages={4204--4215},
  year={2021}
}

@article{oviedo2019fast,
  title={Fast and interpretable classification of small X-ray diffraction datasets using data augmentation and deep neural networks},
  author={Oviedo, F. and Ren, Z. and Sun, S. and others},
  journal={npj Computational Materials},
  volume={5},
  number={1},
  pages={60},
  year={2019}
}

@article{altomare2009advances,
  title={Advances in powder diffraction pattern indexing: N-TREOR09},
  author={Altomare, A. and Campi, G. and Cuocci, C. and Eriksson, L. and Giacovazzo, C. and Moliterni, A. M. and Rizzi, R. and Werner, P.-E.},
  journal={Journal of Applied Crystallography},
  volume={42},
  number={5},
  pages={768--775},
  year={2009}
}

@article{gilmore2004high,
  title={High-throughput powder diffraction. I. A new approach to qualitative and quantitative powder diffraction pattern analysis using full pattern profiles},
  author={Gilmore, C. J. and Barr, G. and Paisley, J.},
  journal={Journal of Applied Crystallography},
  volume={37},
  number={2},
  pages={231--242},
  year={2004}
}

@article{yang2024qwen2,
  title={Qwen2. 5-math technical report: Toward mathematical expert model via self-improvement},
  author={Yang, An and Zhang, Beichen and Hui, Binyuan and Gao, Bofei and Yu, Bowen and Li, Chengpeng and Liu, Dayiheng and Tu, Jianhong and Zhou, Jingren and Lin, Junyang and others},
  journal={arXiv preprint arXiv:2409.12122},
  year={2024}
}

@article{ward2016magpie,
  title={A general-purpose machine learning framework for predicting properties of inorganic materials},
  author={Ward, L. and Agrawal, A. and Choudhary, A. and Wolverton, C.},
  journal={npj Computational Materials},
  volume={2},
  pages={16028},
  year={2016}
}

@article{zaloga2020crystal,
  title={Crystal symmetry classification from powder X-ray diffraction patterns using a convolutional neural network},
  author={Zaloga, A. N. and Stanovov, V. V. and others},
  journal={Materials Today Communications},
  volume={25},
  pages={101662},
  year={2020}
}

@article{vecsei2019neural,
  title={Neural network based classification of crystal symmetries from x-ray diffraction patterns},
  author={Vecsei, P. M. and Choo, K. and Chang, J. and Neupert, T.},
  journal={Physical Review B},
  volume={99},
  number={24},
  pages={245120},
  year={2019}
}

@article{lee2023deep,
  title={A deep learning approach to powder X-ray diffraction pattern analysis: Addressing generalizability and perturbation issues simultaneously},
  author={Lee, B. D. and Lee, J.-W. and Ahn, J. and Kim, S. and Park, W. B. and Sohn, K.-S.},
  journal={Advanced Intelligent Systems},
  volume={5},
  number={9},
  pages={2300140},
  year={2023}
}

@article{banko2021deep,
  title={Deep learning for visualization and novelty detection in large X-ray diffraction datasets},
  author={Banko, L. and Maffettone, P. M. and others},
  journal={npj Computational Materials},
  volume={7},
  number={1},
  pages={104},
  year={2021}
}

@article{cover1967nearest,
  title={Nearest neighbor pattern classification},
  author={Cover, T. M. and Hart, P. E.},
  journal={IEEE Transactions on Information Theory},
  volume={13},
  number={1},
  pages={21--27},
  year={1967}
}

@article{jain2013materialsproject,
  title={The Materials Project: A materials genome approach to accelerating materials innovation},
  author={Jain, A. and Ong, S. P. and Hautier, G. and others},
  journal={APL Materials},
  volume={1},
  number={1},
  pages={011002},
  year={2013}
}

@article{grazulis2012cod,
  title={Crystallography Open Database (COD): an open-access collection of crystal structures},
  author={Grazulis, S. and others},
  journal={Nucleic Acids Research},
  volume={40},
  number={D1},
  pages={D420--D427},
  year={2012}
}

@inproceedings{wu2018nonparametric,
  title={Unsupervised feature learning via non-parametric instance discrimination},
  author={Wu, Z. and Xiong, Y. and Yu, S. X. and Lin, D.},
  booktitle={Proceedings of the IEEE Conference on Computer Vision and Pattern Recognition (CVPR)},
  pages={3733--3742},
  year={2018}
}

@article{tshitoyan2019unsupervised,
  title={Unsupervised word embeddings capture latent knowledge from materials science literature},
  author={Tshitoyan, V. and Dagdelen, J. and Weston, L. and others},
  journal={Nature},
  volume={571},
  pages={95--98},
  year={2019}
}

@article{belsky2002icsd,
  title={New developments in the Inorganic Crystal Structure Database (ICSD)},
  author={Belsky, A. and Hellenbrandt, M. and Karen, V. L. and Luksch, P.},
  journal={Acta Crystallographica Section B},
  volume={58},
  number={3},
  pages={364--369},
  year={2002}
}

@misc{zhang2024matnexus,
  title={MatNexus: a comprehensive text mining and analysis suite for materials discovery. SoftwareX 26, 101654 (2024)},
  author={Zhang, L and Stricker, M},
  year={2024}
}

@article{maffettone2021crystallography,
  title={Crystallography companion agent for high-throughput materials discovery},
  author={Maffettone, Phillip M and Banko, Lars and Cui, Peng and Lysogorskiy, Yury and Little, Marc A and Olds, Daniel and Ludwig, Alfred and Cooper, Andrew I},
  journal={Nature Computational Science},
  volume={1},
  number={4},
  pages={290--297},
  year={2021},
  publisher={Nature Publishing Group US New York}
}

@article{leeman2024challenges,
  title={Challenges in high-throughput inorganic materials prediction and autonomous synthesis},
  author={Leeman, Josh and Liu, Yuhan and Stiles, Joseph and Lee, Scott B and Bhatt, Prajna and Schoop, Leslie M and Palgrave, Robert G},
  journal={PRX Energy},
  volume={3},
  number={1},
  pages={011002},
  year={2024},
  publisher={APS}
}

@article{zhang2025electrocatalyst,
  title={Electrocatalyst discovery through text mining and multi-objective optimization},
  author={Zhang, Lei and Stricker, Markus},
  journal={arXiv preprint arXiv:2502.20860},
  year={2025}
}

@inproceedings{downs2006rruff,
  title={The RRUFF Project: an integrated study of the chemistry, crystallography, Raman and infrared spectroscopy of minerals},
  author={Downs, Robert T},
  booktitle={Program and Abstracts of the 19th General Meeting of the International Mineralogical Association in Kobe, Japan, 2006},
  year={2006}
}

\end{document}